\documentclass{article}
\usepackage{amssymb}

\usepackage{graphicx}
\usepackage{amsmath}
\usepackage{epsfig}
\usepackage{color}

\newenvironment{proof}[1][Proof]{\textbf{#1.} }{\ \rule{0.5em}{0.5em}}

\begin{document}

\title{Rankine-Hugoniot conditions\\ for fluids whose energy depends\\ on space and time  derivatives of density}
\author{S. L. Gavrilyuk$^{\ast }$, H. Gouin\thanks{Aix Marseille Univ, CNRS, IUSTI, UMR 7343, Marseille, France.	
		\newline  E-mails: sergey.gavrilyuk@univ-amu.fr; henri.gouin@univ-amu.fr; henri.gouin@yahoo.fr \qquad   \textbf{Published in Wave Motion 98, 102620 (2020).}}}

\maketitle

\begin{abstract}
By using the Hamilton principle of stationary action, we  derive  the governing equations and  Rankine--Hugoniot  conditions for  continuous media where  the specific energy depends on  the  space and time  density derivatives. The governing system of equations  is a time reversible  dispersive system of conservation laws for the   mass, momentum and energy.   We obtain additional relations to the  Rankine--Hugoniot conditions  coming from the   conservation laws  and  discuss the well-founded of shock wave discontinuities for dispersive systems. 
\end{abstract}

\section{Introduction}
  Ideal shock waves, i.e. surfaces of discontinuities crossed by  mass fluxes, are generally associated with quasilinear hyperbolic systems of conservation laws
  \cite{Courant_1999, Lax_1973, Dafermos, Serre_1996}. Such a moving surface   divides the physical space into two subspaces in which the solution is continuous  but  jumps across the shock. The jump relations  (Rankine--Hugoniot conditions) are derived  from the conservation laws. They relate the normal velocity of the discontinuity surface to the field variables behind and ahead of the shock.      An additional `entropy  inequality'   is usually added to  select admissible shocks \cite{Lax_1973,  Liu}.    For example,  consider the Hopf equation for   unknown function $u(t,x)$ (the choice of the conservative form is postulated {\it a priori}) : 
  \begin{equation}
  \frac{\partial u}{\partial t}+\frac{\partial }{\partial x}\left(\frac{u^2}{2}\right)=0 
  \label{Hopf}
  \end{equation} 
  and the corresponding Riemann problem 
   \begin{equation}
    u(0,x) =\left\{\begin{array}{l}
    u^-\quad {\rm if} \; x<0 \\
    \\
    u^+\quad {\rm if} \; x>0 \\
    \end{array}\right.
    \label{RP}
    \end{equation} 
 If the `entropy' inequality $u^- > u^+$ holds, a unique discontinuous solution is a shock   moving  with  the velocity $D=(u^-+u^+)/2$  :
      \begin{equation}
      u(t,x) =\left\{\begin{array}{l}
     u^-\quad {\rm if} \; x<D \,  t \\
      \\
      u^+\quad {\rm if} \; x>D\,t \\
      \end{array}\right.
      \label{solution_Hopf}
      \end{equation}
  The entropy inequality is usually obtained by the `viscosity method'  : one regularizes the Hopf equation  by the Burgers equation \cite{Serre_1996, Leveque_1992} :
    \begin{equation}
  \frac{\partial u}{\partial t}+\frac{\partial }{\partial x}\left(\frac{u^2}{2}\right)=\nu\,   \frac{\partial^2 u}{\partial x^2},  
  \label{Burgers}
  \end{equation}
  where $\nu \, > 0$ is a small parameter. The condition of existence of  travelling wave solutions for \eqref{Burgers}  joining  the states $u^-$ (respectively $u^+$)  at minus 
  (respectively plus) infinity and having the velocity $D=(u^-+u^+)/2$    yields the inequality  $u^{-}>u^{+}$.  The  travelling wave solution  to \eqref{Burgers}  converges pointwisely  as 
  $\nu \rightarrow 0$  to  the solution \eqref {solution_Hopf}.  However, the viscous regularization does not  always prevent from the existence of discontinuous solutions. For example, in the theory of supersonic boundary layer,  the viscosity is    only present in the direction  transverse to the main  stream, and it  is not sufficient to prevent from the shock formation \cite{Lipatov_Teshukov}.  Analogous results can be found in the theory of long waves in viscous shear flows  down an inclined plane \cite{Chesnokov_Kovtunenko},  and even in compressible flows of a non-viscous but heat-conductive gas \cite{Rozhdestvenskii_Yanenko}. 
  
Usually one thinks that the dispersive regularization excludes the shock formation.  For example, this is the  case of  the Korteweg--de Vries (KdV) equation :
  \begin{equation}
  \frac{\partial u}{\partial t}+u \frac{\partial u}{\partial x}+\nu \frac{\partial^3 u}{\partial x^3} =0  
  \label{KdV}
  \end{equation} 
  where $\nu$ is a small  parameter. The   structure of the solution of     Riemann's problem \eqref{RP} with $u^- >u^+$  to the KdV equation \eqref{KdV} is completely different.  The shocks (first order discontinuities) become  `dispersive shock waves' (DSW)  representing a highly oscillating transition zone joining smoothly the constant states $u^\pm$ \cite{El_Geogjaev_Gurevich_Krylov_1995, Gurevich_Pitaevskii_1974,Gurevich_Krylov_1987, El06, El_2016}.   The leading edge of this DSW approximately represents a half solitary wave propagating  over  the state $u=u^+$ while  the trailing edge of the DSW  corresponds to the small amplitude oscillations near the  state $u^-$. The dispersionless limit $\nu\rightarrow 0$ of \eqref{KdV} is the Whitham system \cite{El_2016}, and not the Hopf equation  \eqref{Hopf}. 
  
  However, let us consider another type of dispersive regularization of \eqref{Hopf} called the  Benjamin--Bona--Mahony (BBM) equation \cite{BBM} : 
    \begin{equation}
  \frac{\partial u}{\partial t}+u \frac{\partial u}{\partial x}-\nu \frac{\partial^3 u}{\partial t\, \partial  x^2} =0, \quad \nu>0.  
  \label{BBM}
  \end{equation} 
A surprising fact is that  the BBM equation admits the  exact stationary {\it discontinuous} solution \cite{El_Royal_Society_2016}. For example, 
 \begin{equation}
u(t,x) =\left\{\begin{array}{l}
-1\quad {\rm if} \; x<0 \\
\\
\ \ \,1\quad {\rm if} \; x>0 \\
\end{array}\right.
\label{solution_BBM}
\end{equation} 
is a stationary solution to \eqref{BBM}. Thus, {\it a priori}, the dispersion does not prevent from   the singular solutions. The solution \eqref{solution_BBM} is an `expansion shock', i.e. it does not verify the Lax stability condition. Such  a shock-like  solution survives  only  a finite time after smoothing of the discontinuity  : the shock structure is conserved but the shock amplitude  is decreasing algebraically  in  time \cite{El_Royal_Society_2016}.
 
One-dimensional shock-like solutions to dispersive equations  connecting  a constant state and  a periodic solution of the governing equations  were recently constructed in \cite{Gavrilyuk_2020} for a  continuum  where the internal energy depended not only on the density but also on its material derivative.  Across the shock considered as a dispersionless limit generalized Rankine--Hugoniot  conditions were satisfied. These conditions are the classical laws for the conservation of mass and  momentum  which are supplemented by an extra condition for the one-sided derivatives of the density coming from the variation of Hamilton's action.     Numerical experiments confirm the relevance of this supplementary  condition \cite{Gavrilyuk_2020}.

One has also to mention the result of \cite{Sprenger_Hoefer_2020} where the  fifth order KdV equation was studied. The dispersionless limit of this equation  is the corresponding  Whitham system. The heteroclinic connection of periodic orbits in the exact equation correspond to the Rankine-Hugoniot conditions for the Whitham system. 

Thus, the difference between the low order dispersive system  \cite{Gavrilyuk_2020} (not admitting heteroclinic connections between periodic orbits) and the higher order dispersive equation (admitting such a heteroclinic connection)  \cite{Sprenger_Hoefer_2020} is the following. In  the first case the generalized shock  relations are obtained from Hamilton's action while in the second case they are obtained from averaged Hamilton's action. 

 We will concentrate here on two classes of low dispersive equations which are Euler--Lagrange equations for    Hamilton's  action  with  a Lagrangian depending not only on the thermodynamic variables but also on their  first spatial and time derivatives.  This is the case  of  the model of fluids endowed with capillarity (the Lagrangian depends on the density gradient)  \cite{Truskinovsky_1982,Casal-Gouin,Dell'Isola-Gouin-Rotoli,Gavrilyuk-Shugrin}  and  the model of  fluids containing gas bubbles \cite{Iordanski, van Wijngaarden,Gavrilyuk-Teshukov, Gavrilyuk-Gouin-Teshukov} (the Lagrangian depends on the material  derivative of density).  Mathematically equivalent models also appear in quantum mechanics  where the nonlinear Schr\"{o}dinger equation is reduced to the equations of capillary fluids via the Madelung transform \cite{Madelung_1927, Carles_2012, Bresch_2019}, and in  the long-wave  theory of  free surface  flows \cite{Gavrilyuk_2004,LannesBOOK_2013, Gavrilyuk_2011} where the equations of motion (Serre--Green--Naghdi equations) have the  form which is equivalent to the equations of bubbly fluids.   We  show that the Hamilton principle implies not only classical Rankine--Hugoniot conditions  for the  mass, momentum and energy, but also additional relations.   The one-dimensional case  where the internal energy depends not only on the density but also on the material derivatives of the  density,  the shock-like transition fronts were already discovered in \cite{Gavrilyuk_2020}. So, it is quite natural to perform the study in the  multi-dimensional case.  For  the continuum where the internal energy  depends on  the density  and density gradient, we hope to present in the future  shock-like solutions in the  case of non-convex  `hydrodynamic' part of the internal energy (in the limit of  vanishing   density gradients the energy is of   van der Waals'  type). 

The technique we use to establish the generalized Rankine--Hugoniot conditions in the multi-dimensional case is related with  the variation of Hamilton's action. To show how it works, we present first a  `toy' system coming from the analytical mechanics.

\begin{figure}[h]
	\begin{center}
		\includegraphics[width=9
		cm]{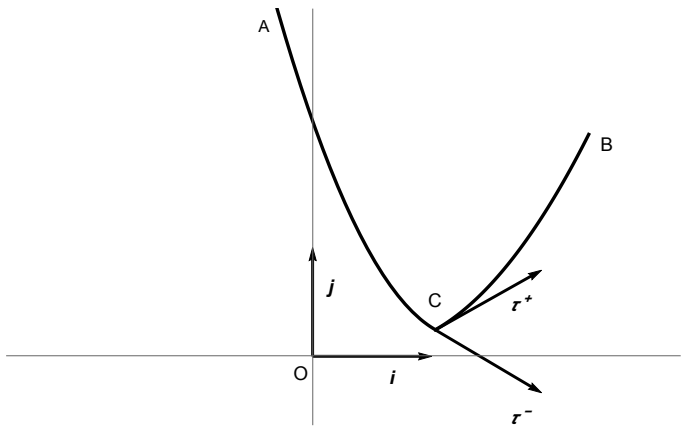}
	\end{center}
	\caption{ A heavy ring  $C$  can freely slide on a heavy curve  of total length $\ell$ and fixed at the end  points $A$ and $B$ in the vertical plane $(O,\,\boldsymbol i,\,\boldsymbol j)$. One needs to find  the position of $C$ as well as the corresponding  equilibrium curve. }
	\label{catenary}
\end{figure}

  A heavy ring  $C$ of mass $m$ can freely slide on a heavy thread of {constant} linear density $\gamma$ having the  total length $\ell$ and fixed at the end  points $A$ and $B$ in the vertical plane $(O,\,\boldsymbol i,\,\boldsymbol j)$ where $\boldsymbol i$ (respectively $\boldsymbol j$) are the horizontal (respectively vertical) unit   vectors (see Figure \ref{catenary}).  We have to determine the position of $C$ as well as the thread form.   To do so, we need to find the extremum of the  system energy  :
\begin{equation*}
W=g\gamma\int_{A}^{B}y(s)ds+mg\,{\boldsymbol j}^T{\boldsymbol C}   
\end{equation*}
submitted to the constraint : 
\begin{equation}
\int_{A}^{B}ds=\ell   \label{length constraint}
\end{equation}
where $\ell$ is a constant length. Here the bold letter ${\boldsymbol C}$ means the vector   {connecting $O$ and $C$},  $g$ is the acceleration of gravity,   $s$ is {the} curvilinear abscissa, and $y(s)$ is the vertical coordinate of the current point of the thread.  Then, the extremum of system energy is associated with :
\begin{equation*}
W^\prime=\int_{A}^{B}n(s)ds+mg\,{\boldsymbol j}^T{\boldsymbol C} 
\end{equation*}
where  $n(s)=g\gamma\, y(s)-{\it\Lambda}$, ${\it \Lambda}$  being a constant Lagrange multiplier associated with {constant length} \eqref{length constraint}. 
Since the ends $A$ and $B$ are fixed, the variation of $W^\prime$  can be written in the form \cite{Gelfand_Fomin_1991, Gouin3} : 
 \begin{eqnarray}
\delta W^\prime &=& -n(C)\left[\boldsymbol{\tau}^T\right] \delta {\boldsymbol C}+mg{\boldsymbol j}^T\delta {\boldsymbol C}\label{catenary1}\\
&+&\left(\int_{A}^C +\int_{C}^B \right)\left(g\,\gamma\,  \boldsymbol{j}^T\left({\boldsymbol I} -{\boldsymbol\tau }{\boldsymbol \tau}^T\right)-\frac{n}{R}\,{\boldsymbol \nu}^T\right)\delta {\boldsymbol{M}}ds=0
\notag
 \end{eqnarray}
where $\delta$ means the variation, $\boldsymbol{\tau}$ and  $\boldsymbol{\nu}$ are the unit tangent and normal vectors to the extremal curve representing the position of the thread, $R$ is the radius of curvature, $\boldsymbol I$ is the identity tensor, square brackets $[...]$ mean  the jump of  $\boldsymbol{\tau}$ at $C$ : $[\boldsymbol{\tau}]=\boldsymbol{\tau}^{+}-\boldsymbol{\tau}^{-}$ (see Figure \ref{catenary}),     ${\boldsymbol M}$ means  the vector connecting $O$ and $M$,  where $M$ is a current point of the curve.     Condition $\delta W'=0$ implies the equation defining  the `broken extremal'  composed of two catenaries given by the solutions of the following equation  \cite{Gouin3} :
\begin{equation*}
g\,\gamma\, \left({\boldsymbol I} -{\boldsymbol\tau }{\boldsymbol \tau}^T\right)\,\boldsymbol{j} -\frac{n}{R}\,{\boldsymbol \nu}  =\boldsymbol 0
\end{equation*}

This equation is  supplemented by two jump conditions at  point $C$ coming from \eqref{catenary1} :
\begin{equation*}
 -n(C)\left[\boldsymbol{\tau}\right]+mg\,{\boldsymbol j}=\boldsymbol 0 
\end{equation*}
Since $n(c)\ne 0$, we finally obtain :
\begin{equation}
\left\{
\begin{array}{l}
\displaystyle 
\left[\boldsymbol{\tau}^T\right]{\boldsymbol i}=0 \\ \\
\displaystyle n(C)\left[\boldsymbol{\tau}^T\right]{\boldsymbol j}-mg =0 
\end{array}\right.\label{Descartes}
\end{equation}
Consequently, the angles between tangent vectors to  the curve  and the horizontal axe  are opposite. The second condition determines the Lagrange multiplier $\it \Lambda$. The conditions \eqref{Descartes}  can be seen as   Rankine--Hugoniot conditions which complement the governing equations and define boundary conditions  for  the  `broken extremal' (Figure \ref{catenary}).

Such a variational technique can be generalized  to the case of continuum mechanics where the variation of  Hamilton's action should naturally be considered in the four-dimensional  physical space-time \cite{Souriau, Gouin1, Gouin-Gavrilyuk}. 

The remainder of this paper is structured as follows. In Section 2 the variation of Hamilton's action with a generic  Lagrangian depending  on the thermodynamic variables and their space-time derivatives is found. The corresponding  Euler--Lagrange equations are specified for the capillary fluids and for the bubbly fluids in Section 3.  The Rankine--Hugoniot conditions for both models are derived in Section 4.  Technical details are given in Appendix.

\section{The Hamilton  action}
\label{The Hamilton  action}

Consider the Hamilton action :
\begin{equation*}
a = \int_{\mathcal W} L\, d\boldsymbol{z}
\end{equation*}
where $\mathcal W$ is a \textit{4-D} domain in space--time. 
The Lagrangian is in the form :
\begin{equation*}
L=\mathcal L\left(\boldsymbol{J},\frac{\partial \boldsymbol{J}}{\partial \boldsymbol{z}}, \eta,\boldsymbol{z}\right)
\label{lagrangian}
\end{equation*}
with    $\boldsymbol{z}= \left(
\begin{array}{c}
t \\
\boldsymbol{x}
\end{array}
\right) \equiv \left\{z^{i}\right\}$,  where $z^0= t$ is the time and $\boldsymbol{x}\equiv \left\{ x^{i}\right\}$, $i=1,2,3$ are the Euler space-variables.    We {write} $\boldsymbol{J}=\left(
\begin{array}{c}
\rho \\
  \rho\boldsymbol{u}
\end{array}
\right)$, where $\rho $ is the  fluid  density,  $\eta$ is the specific entropy (the entropy per unit mass) and $\boldsymbol{u}$ is the fluid velocity. The  \textit{4-D} vector $\boldsymbol{J}$
verifies the  mass conservation :
\begin{equation}
\text{Div}\boldsymbol{J}\equiv \dfrac{\partial \rho }{\partial t}+\text{div}(\rho
\boldsymbol{u})=0  \label{massconstr}
\end{equation}
where $\text{div}$ and $\text{Div}$ are the divergence operators in  the \textit{3-D} physical space  and in  \textit{4-D} physical space-time, respectively.  For conservative motion,  due to \eqref{massconstr}, the equation for the entropy $\eta$ takes the form :
\begin{equation}
\text{Div}\left(\eta \,\boldsymbol{J}\right) \equiv \dfrac{\partial  \rho \eta}{\partial t}+\text{div}(\rho \eta\,
\boldsymbol{u})=0  \label{entropy}
\end{equation}
To calculate the variation of
Hamilton's action,
we consider a \textit{one-parameter family of virtual motions} :
\begin{equation}
{\boldsymbol{z}}=\boldsymbol{\Phi }(\boldsymbol{Z},\varepsilon ) \quad\text{with}\quad\boldsymbol{\Phi }(%
\boldsymbol{Z},0)=\boldsymbol{\varphi}(\boldsymbol{Z})  \label{virtual motions}
\end{equation}
where $\boldsymbol{z}=\boldsymbol{\varphi}  (\boldsymbol{Z})$  represents the real
motion and $\boldsymbol{\Phi }$ is a regular function in the \textit{4-D} reference-space $\mathcal W_0$ of variables
 $\boldsymbol{Z} \equiv \left( Z^{i}\right), (i=0,1,2,3)$ :  $Z^0= \lambda $ is a scalar field (which is not necessarily the time), $\boldsymbol{X}\equiv \left( X^{i}\right), (i=1,2,3)$ are the Lagrange variables. The scalar
$\varepsilon$ is a {real} parameter defined in the vicinity of zero.
   \\ We define  the virtual displacements $ \tilde \delta {\boldsymbol z} (\boldsymbol{Z})$ and the Lagrangian variations $\tilde\delta
\boldsymbol{J(Z)}$  by   the formulas :
\begin{equation*}
\tilde \delta{\boldsymbol z}({\boldsymbol Z})=\left. \frac{\partial
\boldsymbol{\Phi}({\boldsymbol Z},\varepsilon)}{\partial \varepsilon
}\right|_{\varepsilon =0}\label{variation&}
\end{equation*}
\begin{equation*} \tilde\delta\boldsymbol{J(Z)}=\left. \frac{\partial \boldsymbol{J(Z},\varepsilon \boldsymbol{)}}{%
\partial \varepsilon }\right| _{\varepsilon =0}  \label{variation2} 
\end{equation*}
Due to the fact that $\boldsymbol{Z}= 
\boldsymbol\varphi^{-1}\boldsymbol{(z)}$, we can also consider the variations  as
functions of Eulerian variables. Further, we use the  notation 
$\boldsymbol\zeta (\boldsymbol{z})=\tilde \delta\boldsymbol{z}\left(\boldsymbol\varphi^{-1}\boldsymbol(z)\right)$ and we write
$\delta\boldsymbol{J(z)}$ and other quantities without {\it{tilde}} in Eulerian variables: 
 $\delta\boldsymbol{J(z)}=\tilde\delta\boldsymbol{J}\left(\boldsymbol\varphi^{-1}\boldsymbol(z)\right)$. 
The Hamilton  principle assumes
 $\boldsymbol \zeta(\boldsymbol{z})=0$ on the external boundary $\partial\mathcal W $
of $\mathcal W $. \\
{Let $\boldsymbol \zeta^T=\left(\tau, \boldsymbol{\xi}^T\right)$, 
where ${\tau}$ is the scalar part of \textit{4}-vector $\boldsymbol{\zeta}$ associated with time $t$ and \textit{3}-vector $\boldsymbol \xi$ is the part of {\textit{4}-vector $\boldsymbol{\zeta}$} associated with  space-variable $\boldsymbol{x}$}. \footnote{ We  use  the following definitions for  basic  vector  analysis operations. For  vectors $\boldsymbol{a}$ and $\boldsymbol{b}$,   $\boldsymbol{a}%
	^{T }\boldsymbol{b}$ is the scalar product (line vector
	$\boldsymbol{a}^{T}$
	is multiplied by column vector $\boldsymbol{b}$); for the sake of simplicity, we also denote $\boldsymbol{a}%
	^{T }\boldsymbol{a}= \left|\boldsymbol{a}\right|^2$. Tensor $\boldsymbol{a} {%
		\ }\boldsymbol{b}^{T}$ (or $\boldsymbol{a}\otimes \boldsymbol{b}$) is  
	the product of column vector $\boldsymbol{a}$ by line vector $\boldsymbol{b}%
	^{T}$.  Superscript $^T$ denotes the transposition. The divergence of
	a second order tensor ${\boldsymbol A}$ is a covector defined as :
	\begin{equation*}
	\text{Div}\left({\boldsymbol A}\,\boldsymbol{h}\right) =\text{Div}\left({\boldsymbol A}\right) \boldsymbol{h}
	\end{equation*}
	where $\boldsymbol{h}$ is any constant vector field  in the  \textit{4-D} space. In particular,
	one gets for any  \textit{4-D } linear transformation ${\boldsymbol A}$ and any \textit{4-D } vector
	field $\boldsymbol{v}$ :
	\begin{equation*}
	\text{Div}( {\boldsymbol A}\,\boldsymbol{v})=
	(\text{Div }{\boldsymbol A})\,\boldsymbol{v}+ \text{Tr}\left({\boldsymbol A}\,\dfrac{\partial
		\boldsymbol{v}}{\partial \boldsymbol{z}}\right) 
	\end{equation*}
	where $\text{Tr}$ is the trace of a square matrix.   
	Operators $\nabla=\left(\dfrac{\partial}{\partial\boldsymbol{x}}\right)^T$ and $\text{Grad}=\left(\dfrac{\partial}{\partial\boldsymbol{z}}\right)^T$   denote  the     gradients in the \textit{3-D} and \textit{4-D} space, respectively.
	If $f({\boldsymbol A})$ is any scalar function  of ${\boldsymbol A}$, we denote :
	\begin{equation*}
	\nabla_{\boldsymbol A} f = \left(\frac{\partial f}{\partial {\boldsymbol A}}\right)^T  \quad\text{with}\quad\left(\frac{\partial f}{\partial {\boldsymbol A}}\right)_j^i=\left(\frac{\partial f}{\partial {A}_i^j}\right) 
	\end{equation*}
	where  $A^j_i$ are components of ${\boldsymbol A}$ with $i$ being the line index and $j$ being the column index. We   also  denote :
	\begin{equation*}
	df({\boldsymbol A}) = \nabla_{\boldsymbol A} f : d{\boldsymbol A}=\left(\frac{\partial f}{\partial  A_j^i}\right) dA_j^i 
	\end{equation*}
	where repeated indices mean the summation.
	The identity
	matrix and the zero matrix of dimension $k$ are denoted by $\boldsymbol{I}_{k}$
	and $\boldsymbol{O}_{k}$, respectively. The zero vector of dimension $k$ is denoted by $\boldsymbol{0}_{k}$  but, when it has no ambiguity, in the physical \textit{3-D} space,  we simply denote  the zero matrix   by  $\boldsymbol{O}$,  the zero vector  
	by $\boldsymbol{0}$ and  identity tensor by $\boldsymbol{I
	}$.}\\
In   calculations we  use the relation :
\begin{equation}
\delta \boldsymbol{J}=\left( \frac{\partial \boldsymbol \zeta}{\partial \boldsymbol{z}} -\left( \text{Div}\, \boldsymbol \zeta\right)  I_{4}\right) \boldsymbol{J}
\label{moment}
\end{equation}
 (see  Appendix \ref{Appendix 1} for the proof). \\
 Due to \eqref{entropy}, the variation of the specific entropy is zero \cite{Serrin} :
\begin{equation*}
\delta  \eta =0\label{varentro} 
\end{equation*}
It is the reason, we don't always indicate $\eta$ in expression of the Lagrangian, even if the entropy is explicitly presented in the governing equations.\\

The variation of  Hamilton's action 
   is calculated for the family of virtual motions \eqref{virtual motions} :
\begin{equation}
\delta a=\left. \frac{da}{d\varepsilon }\right| _{\varepsilon
=0}=\int_{\mathcal W }\left( \delta L+L\, \text{Div} \, 
\boldsymbol \zeta \right)
d\boldsymbol{z}  \label{var1} 
\end{equation}
and 
\begin{equation*}
\delta L=\frac{\partial L}{\partial \boldsymbol{J}}\ \delta
\boldsymbol{J}+ \text{Tr}\left( \frac{\partial L}{\partial \left(
\dfrac{\partial \boldsymbol{J}}{\partial \boldsymbol{z}}\right) } \,\delta
\left( \frac{\partial \boldsymbol{J}}{\partial \boldsymbol{z}}\right)
\right) +\frac{\partial L}{\partial \boldsymbol{z}}\ \boldsymbol \zeta  
\end{equation*}
To obtain  \eqref{var1} we  use   Jacobi's identity for the generalized \textit{4-D} deformation gradient $\mathcal B$ :
\begin{equation*}\delta\,\text{det}\,\mathcal B = \text{det}\,\mathcal B\  \text{Div} \, 
\boldsymbol \zeta \quad \text{with}\quad \mathcal B = \frac{\partial\boldsymbol{z}}{\partial\boldsymbol{Z}} 
\end{equation*}
To simplify the notation we write :
\begin{equation*}
\mathcal C = \frac{\partial \boldsymbol{J}}{%
	\partial \boldsymbol{z}},\,\  \mathcal A^T =\frac{\partial L}{\partial \left(
	\dfrac{\partial \boldsymbol{J}}{\partial \boldsymbol{z}}\right) }=\left(\nabla_{\mathcal C} L\right)^T\,\ \text{and}\,\,\ \text{Tr}\left(\mathcal A^T \,\delta
\mathcal C\right)
  =\nabla_{\mathcal C} L: \delta \mathcal C 
\end{equation*}
One also has :
\begin{equation*}
\delta \mathcal C=
\delta \left( \frac{\partial \boldsymbol{J}}{\partial \boldsymbol{z}}\right) =\delta\,\left(\frac{\partial \boldsymbol{J} }{\partial \boldsymbol{Z}} \ \frac{\partial \boldsymbol{Z} }{\partial \boldsymbol{z}}\right)=\frac{%
\partial \delta \boldsymbol{J}}{\partial \boldsymbol{z}}-\frac{\partial \boldsymbol{J}}{%
\partial \boldsymbol{z}}\frac{\partial \boldsymbol \zeta}{\partial \boldsymbol{z}} 
\end{equation*}
For the sake of simplicity, in the following  the measure of integration will not be
indicated. Then, we get from (\ref{moment}) and (\ref{var1}) :
\begin{eqnarray*}
	\delta a&=&\int_{\mathcal W }  \frac{\partial L}{\partial
		\boldsymbol{J}}\left( \frac{\partial \boldsymbol \zeta}{\partial \boldsymbol{z}}-\left( \text{Div}\ \boldsymbol \zeta\right) \boldsymbol{I}_{4}\right) \boldsymbol{J}+\text{Tr}\left(\mathcal A^T\left(
	\frac{\partial \delta \boldsymbol{J}}{\partial
		\boldsymbol{z}}-\frac{\partial \boldsymbol{J}}{\partial
		\boldsymbol{z}}\frac{\partial \boldsymbol \zeta}{\partial
		\boldsymbol{z}}\right)  \right) +   \frac{\partial L}{\partial
		\boldsymbol{z}}\ \boldsymbol \zeta+L\, \text{Div}\ \boldsymbol \zeta  
	\end{eqnarray*}
One  has :
\begin{eqnarray*}
	\delta a = \int_{\mathcal W }\text{Tr}\left( \left( \boldsymbol{J}\frac{\partial
		L}{\partial \boldsymbol{J}}-\mathcal A^T\frac{\partial \boldsymbol{J}}{\partial
		\boldsymbol{z}}\right) \frac{\partial {\boldsymbol {\zeta}}}{\partial \boldsymbol{z}}\right) +\left( L-\frac{\partial
		L}{\partial \boldsymbol{J}}\ \boldsymbol{J}\right) \text{Div}\ {\boldsymbol {\zeta}}+
 \text{Tr}\left( \mathcal A^T\frac{\partial \delta \boldsymbol{J}}{\partial
			\boldsymbol{z}}\right) +\frac{\partial L}{\partial \boldsymbol{z}}\
		{ {\boldsymbol {\zeta}}}\\
	=\int_{\mathcal W } \text{Tr}\left( \left( \boldsymbol{J}\frac{\partial
		L}{\partial \boldsymbol{J}} - \mathcal A^T\frac{\partial \boldsymbol{J}}{\partial
		\boldsymbol{z}}\right) \frac{ {\boldsymbol {\partial\zeta}}}{\partial \boldsymbol{z}}\right)  + \left( L-
	\frac{\partial L}{\partial \boldsymbol{J}}\ \boldsymbol{J}\right) \text{Div}\
	 {\boldsymbol {\zeta}}  
	 +  \text{Div}\left( \mathcal A^T\delta \boldsymbol{J}\right) -\text{Div}\left(
	\mathcal A^T\right) \delta \boldsymbol{J}+\frac{\partial L}{\partial
		\boldsymbol{z}}\ {\boldsymbol {\zeta}} 
	\end{eqnarray*}
and finally :
	\begin{eqnarray*}
	\delta a &=&\int_{\mathcal W }\text{Tr}\left( \left( \boldsymbol{J}\frac{\partial
		L}{\partial \boldsymbol{J}}-\mathcal A^T\frac{\partial \boldsymbol{J}}{\partial
		\boldsymbol{z}}\right) \frac{\partial \boldsymbol \zeta}{\partial \boldsymbol{z}}\right) \boldsymbol{+}\left(
	L-\frac{\partial L}{\partial \boldsymbol{J}}\ \boldsymbol{J}\right) \text{Div}\
	\boldsymbol \zeta+\text{Div}\left( \mathcal A^T\delta \boldsymbol{J}\right)\\
	&-& \int_{\mathcal W }\text{Div}\left( \mathcal A^T\right) \left( \frac{\partial {\boldsymbol{\zeta}}}{\partial \boldsymbol{z}}-\left( \text{Div}\ {{\boldsymbol{\zeta}}}\right) \boldsymbol{I}_{4}\right) \boldsymbol{J}+\frac{\partial
		L}{\partial \boldsymbol{z}}\, {\boldsymbol{\zeta}} \\
	&=&\int_{\mathcal W }\text{Tr}\left( \left( \boldsymbol{J}\frac{\partial
		L}{\partial \boldsymbol{J}}-\mathcal A^T\frac{\partial \boldsymbol{J}}{\partial
		\boldsymbol{z}}-\boldsymbol{J}\,\text{Div}\left(\mathcal A^T\right) \right)
	\frac{\partial {\boldsymbol{\zeta}}}{\partial \boldsymbol{z}}
	\right)  \\
	&+&\int_{\mathcal W }\left( L-\frac{\partial L}{\partial \boldsymbol{J}}\
	\boldsymbol{J}+\left( \text{Div} \mathcal A^T\right) \boldsymbol{J}\right) \text{Div}\
	{{\boldsymbol{\zeta}}}+\text{Div}\left(\mathcal A^T\delta \boldsymbol{J}\right)
	+\frac{\partial L}{\partial \boldsymbol{z}}\ \boldsymbol \zeta \\
		&=&\int_{\mathcal W }\text{Div}\left( \left( \boldsymbol{J}\frac{\partial
			L}{\partial \boldsymbol{J}}-\mathcal A^T\frac{\partial \boldsymbol{J}}{\partial
			\boldsymbol{z}}-\boldsymbol{J}\
		\text{Div}\left(\mathcal A^T\right) \right)   {{\boldsymbol{\zeta}}} +\mathcal A^T\delta \boldsymbol{J}%
		\right)  \\
		&-&\int_{\mathcal W }\text{Div}\left( \boldsymbol{J}\frac{\partial L}{\partial \boldsymbol{J}}-\mathcal A^T\frac{%
			\partial \boldsymbol{J}}{\partial \boldsymbol{z}}-\boldsymbol{J} \,\text{Div}\left(
		\mathcal A^T\right) \right)  \boldsymbol \zeta \\
		&+&\int_{\mathcal W }\text{Div}\left( \left( L-\frac{\partial L}{\partial
			\boldsymbol{J}}\boldsymbol{J}+\left( \text{Div}\mathcal A^T\right) \boldsymbol{J}\right)  
		{{\boldsymbol{\zeta}}}\right)\\ &-&\int_{\mathcal W }\text{Grad}\left( L- \frac{\partial
			L}{\partial \boldsymbol{J}}\boldsymbol{J}+\left( \text{Div}\mathcal A^T\right)
		\boldsymbol{J}\right) ^T  \boldsymbol \zeta+\frac{\partial
			L}{\partial \boldsymbol{z}}\ {{\boldsymbol{\zeta}}}
	\end{eqnarray*}
Let us denote  :
\begin{equation*}
\boldsymbol{K}^T=\frac{\delta L}{\delta \boldsymbol{J}}\equiv \frac{\partial
L}{\partial \boldsymbol{J}}- \text{Div}\left( \mathcal A^T\right) 
\end{equation*}
then :
	\begin{eqnarray*}
		\delta a &=&\int_{\mathcal W }\left\{ \frac{\partial L}{\partial \boldsymbol{z}}%
		-\text{Div}\left( \boldsymbol{J} \boldsymbol{K}^T-\mathcal A^T\frac{\partial \boldsymbol{J}}{%
			\partial \boldsymbol{z}}+p\,%
		\boldsymbol{I}_{4}\right) \right\} \ {{\boldsymbol{\zeta}}} 
		\label{actionvar}\\
		&+&\int_{\partial \mathcal W }{\boldsymbol{N}^T}\left( \boldsymbol{J}\
		\boldsymbol{K}
		^T-\mathcal A^T\frac{\partial \boldsymbol{J}}{\partial \boldsymbol{z}}+p\,\boldsymbol{I}_{4}\right){{\boldsymbol{\zeta}}}+{\boldsymbol{N}^T}\mathcal A^T\delta \boldsymbol{J} \notag
	\end{eqnarray*}
where 
\begin{equation}
p= L-\boldsymbol{K}^T  \boldsymbol{J}\label{pressure}
\end{equation}  
and 
 $\boldsymbol{N}^T$ is a \textit{4-D} co-vector  canceling the tangent vectors to $\partial \mathcal W$.
 \\
 
 {Due to the fact, the virtual displacement can be considered as null in the vicinity of the boundary $\partial \mathcal W $, the Hamilton principle simply  writes :
 \begin{eqnarray*}
 \delta a &=&\int_{\mathcal W }\left\{ \frac{\partial L}{\partial \boldsymbol{z}}%
 -\text{Div}\left( \boldsymbol{J} \boldsymbol{K}^T-\mathcal A^T\frac{\partial \boldsymbol{J}}{%
 	\partial \boldsymbol{z}}+p\,%
 \boldsymbol{I}_{4}\right) \right\} \ {{\boldsymbol{\zeta}}} =0
 \label{HP} 
 \end{eqnarray*}
and we get the
equations of motions in a conservative form  \cite{Gavrilyuk-Gouin} : 
\begin{equation}
\text{ Div}\left( \boldsymbol{J}\,\boldsymbol{K}^T-\mathcal A^T\frac{\partial \boldsymbol{J}}{%
	\partial \boldsymbol{z}}+p\,%
\boldsymbol{I}_{4}\right)  = \frac{\partial L}{\partial \boldsymbol{z}} 
\label{Motion equations} 
\end{equation}
where term $\partial L/\partial \boldsymbol z$ is associated with the external body forces.\\
 
 Let the motion be discontinuous on a \textit{3-D} surface $\it\it \Sigma$ with normal $ \boldsymbol N$  (see Figure \ref{shock}). The virtual displacement can be considered as null in the vicinity of the boundary $\partial \mathcal W $, and equations of motions \eqref{Motion equations} being satisfied, the variation of the Hamilton action is reduced to :
  
\begin{equation}
\delta a  =  \int_{\it \it \Sigma }\left[ {\boldsymbol{N}^T}\left( \boldsymbol{J}\boldsymbol{K}%
^T-A^T\frac{\partial \boldsymbol{J}}{\partial \boldsymbol{z}}+p\,  \boldsymbol{I}_{4}\right)   {{\boldsymbol{\zeta}}}+{\boldsymbol{N}^T}A^T\delta \boldsymbol{J}\right] \label{actionvar1}
\end{equation}
where the  brackets $\left[ \,\,
\right] $ mean the jumps of discontinuous quantities across $\it \it \Sigma$. 
\begin{figure}[h]
	\begin{center}
		\includegraphics[width=7
		cm]{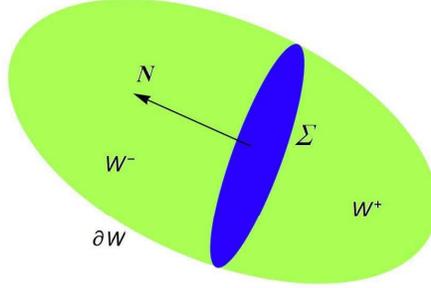}
	\end{center}
	\caption{The shock surface $\it \it \Sigma$ in $\mathcal W$:  The    shock surface   $\it \it \Sigma$  divides the  space-time domain $\mathcal W$ into two space-time domains $\mathcal W^-$ and $\mathcal W^+$ behind  and ahead of the shock in which the solution is continuous  but across  the shock.  }
	\label{shock}
\end{figure}

 In the next section, we explicit the governing equation \eqref{Motion equations}  for two  cases : second gradient fluids and bubbly fluids.  Then, we present the surface integral \eqref{actionvar1}  for these two specific cases.

 \section{Particular cases}
 {\subsection{Capillary fluids} Without body forces, the Lagrangian is of the form :  
\begin{equation*}
L=\frac{1}{2}\,\rho \, \left|\boldsymbol{u}\right|^2 -\rho\, \alpha\left( \rho , \frac{\partial\rho}{\partial\boldsymbol x}, \eta\right)  \equiv \frac{1}{2\,\rho} \, \left|\boldsymbol{j}\right|^2  -\rho\, \alpha\left( \rho , \frac{\partial\rho}{\partial\boldsymbol x}, \eta\right) 
\end{equation*}
Then 
\begin{equation}
\frac{\partial \boldsymbol{J}}{\partial \boldsymbol{z}}=\left(
\begin{array}{ll}
\;\dfrac{\partial \rho }{\partial t} & \;\dfrac{\partial \rho }{\partial
	\boldsymbol{x}} \\
\begin{array}{l}
\\
\dfrac{\partial \boldsymbol{j}}{\partial t} 
\end{array}
&
\begin{array}{l}
\\
\dfrac{\partial \boldsymbol{j}}{\partial \boldsymbol{x}}
\end{array}
\end{array}
\right), \quad 
\frac{\partial L}{\partial \boldsymbol{J}}=\left( -\frac{1}{2}\,\left|\boldsymbol{u}\right|^2  -\alpha-\rho\,\dfrac{\partial \alpha}{\partial \rho } 
,\ \boldsymbol{u}^T \right) 
\label{Partial J}
\end{equation}
We deduce :
\begin{equation}
\mathcal A^T=\frac{\partial L}{\partial \left( \dfrac{\partial \boldsymbol{J}}{\partial
		\boldsymbol{z}}\right) }=\left(
\begin{array}{ll}
\quad\quad\quad 0 & \;\boldsymbol{0}_3^T
\\
\begin{array}{l}
\\
-\rho\,\dfrac{\partial\alpha}{\partial\left(\dfrac{\partial\rho}{\partial\boldsymbol x}\right)} 
\end{array}
&
\begin{array}{l}
\\
O_{3}
\end{array}
\end{array}
\right) 
\label{formulae2} 
\end{equation}
and get :
\begin{equation*}
\text {Div}\,(\mathcal A^T)=-\left( \text{div}\left\{\rho\,\dfrac{\partial\alpha}{\partial\left(\dfrac{\partial\rho}{\partial\boldsymbol x}\right)}\right\},\;\boldsymbol{0}^T_3\right) 
\end{equation*}
\begin{equation}
	\boldsymbol{K}^T =\frac{\partial L}{\partial \boldsymbol{J}}-\text {Div}\,(\mathcal A^T)= \left( -\frac{1}{2}\left|\boldsymbol{u}\right|^2-\alpha-\rho\,\dfrac{\partial \alpha}{\partial\rho
	}+\text{div}\left\{\rho\,\dfrac{\partial\alpha}{\partial\left(\dfrac{\partial\rho}{\partial\boldsymbol x}\right)}\right\},\ 
	\boldsymbol{u}^T \right)  \label{K^T}
\end{equation}
 The pressure $p$ deduced from \eqref{pressure} is defined as : 
\begin{equation*}
	p\equiv \rho^2 \,  \dfrac{\partial \alpha}{\partial\rho }-\rho\ \text{div}\left(\rho\,\dfrac{\partial\alpha}{\partial\left(\dfrac{\partial\rho}{\partial\boldsymbol x}\right)}\right)  \label{pressure capillary}
\end{equation*}
  Nevertheless, $p$ is not   only  a function  of
density as in the case of barotropic fluids: it depends also on
the density gradient. 
Additively :
\begin{equation*}
	\mathcal A^T\frac{\partial \boldsymbol{J}}{\partial \boldsymbol{z}}=\left(
	\begin{array}{ll}
		\qquad \quad 0, & \qquad\quad 
		\boldsymbol{0}_3^T \\
		\begin{array}{l}
			\\
			-\rho\,\dfrac{\partial\alpha}{\partial\left(\dfrac{\partial\rho}{\partial\boldsymbol x}\right)} \dfrac{\partial \rho }{\partial t}, 	 
		\end{array}
		&
		\begin{array}{l}
			\\
			-\rho\,\dfrac{\partial\alpha}{\partial\left(\dfrac{\partial\rho}{\partial\boldsymbol x}\right)}\dfrac{\partial \rho }{\partial
				\boldsymbol{x}} 
		\end{array}
	\end{array}
	\right) 
\end{equation*}
\begin{equation*}
	\boldsymbol{J} \boldsymbol{K} ^T =\left(
	\begin{array}{ll}\quad
		-\dfrac{1 }{2} \rho\,\left|\boldsymbol{u}\right|^2 -  \rho\,\alpha  -p,& \;\  \rho\,	\boldsymbol{u}^T  \\
		\begin{array}{l}
			\\
			\left(-\dfrac{1 }{2} \rho\,	\left|\boldsymbol{u}\right|^2 - \rho\, \alpha -p \right) \boldsymbol{u},
		\end{array}
		&
		\begin{array}{l}
			\\
			\rho\,\boldsymbol{u}	\boldsymbol{u}^T 
		\end{array}
	\end{array}
	\right)  
\end{equation*}
Equation \eqref{Motion equations} writes :
\begin{equation}
	{\text {Div}} \left(
	\begin{array}{ll}
		\qquad\qquad  e, & \qquad \qquad -\rho\,	\boldsymbol{u}^T\\
		\begin{array}{l}
			\\
		\left(e + p \right) \boldsymbol{u}-\rho\,\dfrac{\partial\rho}{\partial t}\dfrac{\partial\alpha}{\partial\left(\dfrac{\partial\rho}{\partial\boldsymbol x}\right)},
		\end{array}
		&
		\begin{array}{l}
			\\
			-\rho\,\boldsymbol{u}	\boldsymbol{u}^T-\rho\,\dfrac{\partial\alpha}{\partial\left(\dfrac{\partial\rho}{\partial\boldsymbol x}\right)} \dfrac{\partial \rho }{\partial \boldsymbol{x}}-p\, \boldsymbol{I}_{3}
		\end{array}
	\end{array}
	\right) = \boldsymbol{0}^T_4   \label{motiongrad}
\end{equation}
where
we denote by 
\begin{equation*}
	e= \rho\,\left(\frac{1}{2}\,\left|\boldsymbol{u}\right|^2+\alpha\right)
\end{equation*}
the total energy per unit volume of the fluid; the first component of Eq. \eqref{motiongrad} yields the {\it equation of energy} :
\begin{equation}
 \frac{\partial	e}{\partial t}+ {\text div }\left(\left(e+p\right)\boldsymbol  u-\rho\,\dfrac{\partial\rho}{\partial t}\dfrac{\partial\alpha}{\partial\left(\dfrac{\partial\rho}{\partial\boldsymbol x}\right)}\right)=0 \label{energy capillary}
\end{equation}
The additive term  $    \rho\,\dfrac{\partial\rho}{\partial t}\dfrac{\partial\alpha}{\partial\left(\dfrac{\partial\rho}{\partial\boldsymbol x}\right)}$ is a flux of energy corresponding to the  interstitial working \cite{Casal-Gouin}.\\
The three other components of  Eq. \eqref{motiongrad} yield the {\it equations of motion} :
\begin{equation}
	\frac{\partial	\rho\, \boldsymbol{u}^T}{\partial t}+\text{div}  \left(\rho\,\boldsymbol{u}\otimes \boldsymbol{u}+p\,\boldsymbol I
+\rho\,\dfrac{\partial\alpha}{\partial\left(\dfrac{\partial\rho}{\partial\boldsymbol x}\right)}\dfrac{\partial\rho}{\partial \boldsymbol  x}\right)= \boldsymbol 0^T \label{motion capillary}
\end{equation}
  Equations \eqref{energy capillary} and \eqref{motion capillary} represent the development of \eqref{Motion equations} for capillary fluids.

 \subsection{Bubbly fluids} Without body forces, the Lagrangian  is of the form :
\begin{equation*}
 L=\frac{1}{2}\,\rho \, \left|\boldsymbol{u}\right|^2-\rho\, w\left( \rho,\dot\rho,  \eta \right) \equiv \frac{1}{2\,\rho} \, \left|\boldsymbol{j}\right|^2  -\rho\, w\left( \rho,\dot\rho,  \eta\right) 
\end{equation*}
 where 
 \begin{equation*}
 \overset{\centerdot}{\rho  }=%
 \dfrac{d\rho}{dt}=\dfrac{\;\partial\rho}{\partial
 	t}+\frac{\partial \rho}{\partial \boldsymbol{x}}\,{\boldsymbol{u}} \equiv \dfrac{\partial\rho}{\partial
 	t}+\frac{1}{\rho} \,\frac{\partial \rho}{\partial \boldsymbol{x}}\,{\boldsymbol{j}}
 \label{ll}
 \end{equation*}
Such a Lagrangian appears in the study of wave propagation for
 shallow water flows with dispersion and bubbly flows (a complete
 discussion of these models is given in \cite{Gavrilyuk-Teshukov}). Then    \eqref{Partial J} can be explicitly written as  :
  \begin{equation*}
 \frac{\partial L}{\partial \boldsymbol{J}}=\left( -\frac{1}{2}\,\left|\boldsymbol{u}\right|^2 -w-\rho\,\dfrac{\partial w}{\partial \rho }+
  \dfrac{\partial w}{\partial \overset{\centerdot }{\rho }}\frac{\partial\rho}{\partial \boldsymbol{x}}\,  \boldsymbol{u },\,\ \boldsymbol{u}^T-\dfrac{%
  	\partial w}{\partial \overset{\centerdot }{\rho }}\frac{\partial\rho}{\partial \boldsymbol{x}}\right) 
  \end{equation*}
  We deduce :
  \begin{equation}
  \mathcal A^T=\frac{\partial L}{\partial \left( \dfrac{\partial \boldsymbol{J}}{\partial
  		\boldsymbol{z}}\right) }=\left(
  \begin{array}{ll}
  \ \;-\rho\,\dfrac{\partial w}{\partial \overset{\centerdot }{\rho }} & \;\boldsymbol{0}_3^T
  \\
  \begin{array}{l}
  \\
  -\rho\,\dfrac{\partial w}{\partial \overset{\centerdot }{\rho }}\,\boldsymbol{u}
  \end{array}
  &
  \begin{array}{l}
  \\
  O_{3}
  \end{array}
  \end{array}
  \right) 
  \label{formulae} 
  \end{equation}
 Due to \eqref{massconstr} we obtain :
  \begin{equation*}
  \text {Div}\,(\mathcal A^T)=\left( -\dfrac{\partial }{\partial t}\left( \rho\,\dfrac{\partial w}{%
  	\partial \overset{\centerdot }{\rho }}\right) -\text{div}\left(\rho\, \dfrac{\partial w}{%
  	\partial \overset{\centerdot }{\rho }}\boldsymbol{u}\right),\;\boldsymbol{0}^T_3\right)\equiv\left( -\rho\,\dfrac{d }{dt}\left( \dfrac{\partial w}{%
  	\partial \overset{\centerdot }{\rho }}\right) ,\;\boldsymbol{0}^T_3\right) 
  \end{equation*}
\begin{eqnarray*}
\boldsymbol{K}^T  &=&\frac{\partial L}{\partial \boldsymbol{J}}-\text {Div}\,(\mathcal A^T)\\
&=&\left( -\frac{1}{2}\,\left|\boldsymbol{u}\right|^2-w-\rho\,\dfrac{\partial w}{\partial\rho
}+ \dfrac{\partial w}{\partial
	\overset{\centerdot }{\rho }}  \frac{\partial\rho}{\partial \boldsymbol{x}}\,\boldsymbol{u}+\rho\,\dfrac{d }{dt}\left( \dfrac{\partial w}{%
	\partial \overset{\centerdot }{\rho }}\right) ,\ \
\boldsymbol{u}^T- \dfrac{\partial w}{\partial \overset{%
		\centerdot }{\rho }}\frac{\partial\rho}{\partial \boldsymbol{x}} \right)\label{Kbis}
  \end{eqnarray*}
  The pressure $p$ obtained from \eqref{pressure} is defined as :
  \begin{equation*}
  p=
  \rho^2 \left( \dfrac{\partial w}{\partial\rho }-\dfrac{d}{%
  	dt}\left( \dfrac{\partial w}{\partial \overset{\centerdot }{\rho }}%
  \right)\right) \label{pressure bubbly}
  \end{equation*}
Nevertheless  pressure $p$  depends also on the material  derivatives of the density.  
  Additively,
  \begin{equation*}
  \mathcal A^T\frac{\partial \boldsymbol{J}}{\partial \boldsymbol{z}}=\left(
  \begin{array}{ll}
  \;-\rho\,\dfrac{\partial w}{\partial \overset{\centerdot }{\rho }}\dfrac{\partial \rho }{\partial t}, & \;-\rho\,\dfrac{\partial w}{\partial \overset{\centerdot }{\rho }}\dfrac{\partial \rho }{\partial
  	\boldsymbol{x}} \\
  \begin{array}{l}
  \\
  -\rho\,\dfrac{\partial w}{\partial \overset{\centerdot }{\rho }}\dfrac{\partial \rho }{\partial t} 	\boldsymbol{u},
  \end{array}
  &
  \begin{array}{l}
  \\
  -\rho\,\dfrac{\partial w}{\partial \overset{\centerdot }{\rho }}	\boldsymbol{u} \dfrac{\partial \rho }{\partial
  	\boldsymbol{x}} 
  \end{array}
  \end{array}
  \right)  
  \end{equation*}
  \begin{equation*}
  \boldsymbol{J} \boldsymbol{K} ^T =\left(
  \begin{array}{ll}
  -\dfrac{1 }{2} \rho\,	\left|\boldsymbol{u}\right|^2 - \rho\, w  -p+ \rho\,\dfrac{\partial w}{\partial \overset{\centerdot }{\rho }}\dfrac{\partial \rho }{\partial 	\boldsymbol{x}} 	\boldsymbol{u},& \;\  \rho\,	\boldsymbol{u}^T- \rho\,\dfrac{\partial w}{\partial \overset{\centerdot }{\rho }}\dfrac{\partial \rho }{\partial 	\boldsymbol{x}} \\
  \begin{array}{l}
  \\
  \left(-\dfrac{1 }{2} \rho\,	\left|\boldsymbol{u}\right|^2- \rho\, w -p+\rho\, \dfrac{\partial w}{\partial \overset{\centerdot }{\rho }}\dfrac{\partial \rho }{\partial 	\boldsymbol{x}} 	\boldsymbol{u}\right) \boldsymbol{u},
  \end{array}
  &
  \begin{array}{l}
  \\
  \rho\,\boldsymbol{u}	\boldsymbol{u}^T-\rho\, \dfrac{\partial w}{\partial \overset{\centerdot }{\rho }}\boldsymbol{u}\dfrac{\partial \rho }{\partial 	\boldsymbol{x}} 
  \end{array}
  \end{array}
  \right)   
  \end{equation*}
 Equation \eqref{Motion equations} writes :
  \begin{equation}
  {\text {Div}} \left(
  \begin{array}{ll}
  \quad \dfrac{1 }{2} \rho\,	\left|\boldsymbol{u}\right|^2 +  \rho\,w-  \rho\,\dfrac{\partial w}{\partial \overset{\centerdot }{\rho }} \overset{\centerdot }{\rho }, & \;\  -\rho\,	\boldsymbol{u}^T\\
  \begin{array}{l}
  \\
  \left(\dfrac{1 }{2} \rho\,	\left|\boldsymbol{u}\right|^2 +  \rho\,w- \rho\, \dfrac{\partial w}{\partial \overset{\centerdot }{\rho }} \overset{\centerdot }{\rho } + p\right) \boldsymbol{u},
  \end{array}
  &
  \begin{array}{l}
  \\
  -\rho\,\boldsymbol{u}	\boldsymbol{u}^T-p\, \boldsymbol{I}_{3}
  \end{array}
  \end{array}
  \right) = \boldsymbol{0}^T_4   \label{motionbub}
  \end{equation}
We denote by 
  \begin{equation*}
  e = \rho\,\left(\frac{1}{2}\,\left|\boldsymbol{u}\right|^2+w-\overset{\centerdot }{\rho }\,\frac{\partial w}{\partial \overset{\centerdot }{\rho }}\right)  
  \end{equation*}
the total energy per unit volume of the fluid; the first component of Eq. \eqref{motionbub} yields the {\it equation of energy} :
  \begin{equation}
   \frac{\partial e }{\partial t}+ {\text div }\left(\left(e+p\right)\boldsymbol  u\right)  =0 \label{energy bubbly}
  \end{equation}
  The three other components of  Eq. \eqref{motionbub} yield the {\it equations of motion} :
  \begin{equation}
  \frac{\partial \rho\boldsymbol u^T}{\partial t}+\text{div}(\rho\boldsymbol u\otimes\boldsymbol u+ p\,I)={\bf 0}^T\label{motion bubbly}
  \end{equation}
  Equations \eqref{energy bubbly} and \eqref{motion bubbly} represent the development of \eqref{Motion equations} for bubbly fluids.
  
 \section{ Rankine--Hugoniot conditions}
We   represent $\it \it \Sigma$ in the form ${\it \Sigma}=[t_0,t_1]\times S_t$ where $S_t$ is  a time  dependent   surface.  Then, $\boldsymbol{N}^T=(-D_{n},\boldsymbol{n}^T)$, where $D_{n}$ denotes the
 normal surface velocity of $S_t$  and $\boldsymbol{n}$ is the  unit  normal vector to  $S_t$.} The mass conservation law \eqref{massconstr} yields   the relation  :
\begin{equation*}
\left[\boldsymbol N^T \boldsymbol J\right]=\left[\rho\left(\boldsymbol n^T\boldsymbol{u}-D_n\right)\right]= [\rho\, v]=0 \qquad \text{with}\,\ v = \boldsymbol n^T\boldsymbol{u}-D_n
\end{equation*}
{As seen in Appendix 2,} in the two particular cases,  the term $ \boldsymbol {N}^T \mathcal A^T\delta \boldsymbol {J}$ can be   written as  :
\begin{equation*}
\boldsymbol {N}^T \mathcal A^T\delta \boldsymbol {J}=b\, \left(\text{div}\,\boldsymbol{\xi}-\frac{\partial \tau}{\partial \boldsymbol{x}}\boldsymbol{u}\right) 
\end{equation*}
where $b$ is the scalar which is given in explicit form for capillary fluids by \eqref{bcapillary} and for bubbly fluids by \eqref{bbubbly}  \footnote{We can also remark (see \cite{Gouin1})  that $\displaystyle \delta \boldsymbol{F}   
	= \ \left( {\partial {  {\boldsymbol \xi}}\over \partial
		{\boldsymbol x}}  - {\boldsymbol u} \, {\partial {\tau}\over \partial {\boldsymbol x}}\right)\boldsymbol{F}\  $  where $\displaystyle\boldsymbol{F} = \frac{\partial \boldsymbol x}{\partial \boldsymbol X}$ and $ \displaystyle\text{Tr}\left({\partial {  {\boldsymbol \xi}}\over \partial
		{\boldsymbol x}}  - {\boldsymbol u} \, {\partial {\tau}\over \partial {\boldsymbol x}}\right)=\text{div}\,\boldsymbol{\xi}-\frac{\partial \tau}{\partial \boldsymbol{x}}\boldsymbol{u}$.  Consequently,  the variation of  $\boldsymbol {N}^T \mathcal A^T\delta \boldsymbol {J}$ is associated with the change of volume. }. 
\\

\noindent We study the case when $\it \it \Sigma$ is a shock surface and 
consequently   $\rho\,v=\boldsymbol{N}^T\boldsymbol{J}\neq0$}. \\

\noindent  Equation \eqref{Motion equations} being verified, we obtain for all vector field $\boldsymbol \zeta$  the variation of $\delta a$ {in the  form  \eqref{actionvar1}. The surface integral \eqref{actionvar1} will be presented in separable form 
in terms of virtual displacements and  their normal derivatives along the \textit{3-D} manifold $\it \Sigma$ \cite{Schwartz, Gouin2}.\\ 
Using Lemmas from Appendix \ref{Technical lemmas}, we obtain :
	\begin{eqnarray*}
		\boldsymbol {N}^T \mathcal A^T\delta \boldsymbol {J}  &=&b\, \left(\text{div}\,\boldsymbol{\xi}-\frac{\partial \tau}{\partial \boldsymbol{x}}\boldsymbol{u}\right) = b\,  \text{div}( \boldsymbol \xi -\tau  \boldsymbol u)+\tau\, b\,  \text{div}(\boldsymbol u) \\
		&=& -\boldsymbol {\it \Theta}^T\boldsymbol{\zeta}-b\,\boldsymbol{n}^T \boldsymbol u\,\frac{d\tau}{dn}+ b\,\frac{d{\xi}_n}{dn}+\boldsymbol {n}^T\, {\rm rot}\,\left(\boldsymbol {n}\times b\,\left(\boldsymbol \xi -\tau\boldsymbol u\right)\right),
	\end{eqnarray*}
with 
\begin{equation}
\boldsymbol {\it \Theta} =\left(
\begin{array}{c}
-b\, H  \boldsymbol{n}^T \boldsymbol u-\text{div}_{tg}\,(b\,\boldsymbol u) \\
\\
b\, H  \boldsymbol{n}   + \nabla_{tg} b+\displaystyle b\frac{\partial \boldsymbol n }{\partial \boldsymbol x}\boldsymbol n
\end{array}
\right).
\label{Theta}
\end{equation}
Here  we use the notations  $\displaystyle\nabla_{tg}^T b=\dfrac{\partial b}{\partial
	\boldsymbol{x}}\left( \boldsymbol{I}-\boldsymbol{n}\boldsymbol{n}^T\right) $ and  $\displaystyle\text{div}_{tg}(b\mathbf u)=\text{div}\,\left(b\,\boldsymbol u\right)-\boldsymbol{n}^T\,\frac{\partial b\,\boldsymbol u}{\partial \boldsymbol x}\, \boldsymbol{n}$, where index `{\it tg}' means the tangential gradient and  tangential divergence operators to $S_t$. Also, $H=-\text{div}(\boldsymbol n)$ is the sum of  principal curvatures of $S_t$.
From \eqref{actionvar1}, one obtains :
\begin{eqnarray*} 
\delta a &=&\int_{\it \Sigma }  \left[\left\{\boldsymbol{N}^T\left( \boldsymbol{J}\boldsymbol{K}%
^T-\mathcal A^T\frac{\partial \boldsymbol{J}}{\partial \boldsymbol{z}}+p\,  \boldsymbol{I}_{4}\right)-\boldsymbol{\it \Theta}^T\right\}\boldsymbol\zeta \right]\\
 &-&\int_{\it \Sigma }\left[ b\,\boldsymbol{n}^T \boldsymbol u\,\frac{d\tau}{dn}- b\,\frac{d{\xi}_n}{dn}\right]+\int_{t_0}^{t_1}\int_{C_t}\left[b \, \left\{\boldsymbol {n}\times  \left(\boldsymbol \xi -\tau\boldsymbol u\right)\right\}^T \boldsymbol {t}\right]
\end{eqnarray*} 
Here $C_t$ denotes the boundary of $S_t$ ($\it \Sigma = [t_0, t_1]\times S_t$), and $\boldsymbol t$ is the oriented unit tangent vector  to $C_t$. Since we are looking for shock relations, the virtual displacements are vector fields with compact support on $S_t$, and  the integral on $C_t$ is vanishing :
\begin{equation} 
	\delta a =\int_{\it \Sigma }  \left[\left\{\boldsymbol{N}^T\left( \boldsymbol{J}\boldsymbol{K}%
	^T-\mathcal A^T\frac{\partial \boldsymbol{J}}{\partial \boldsymbol{z}}+p\,  \boldsymbol{I}_{4}\right)-\boldsymbol{\it \Theta}^T\right\}\boldsymbol\zeta 
	- b\,\boldsymbol{n}^T \boldsymbol u\,\frac{d\tau}{dn}+ b\,\frac{d{\xi}_n}{dn}\right] 
	\label{key4}
\end{equation}  
This expression of $\delta a$ is in  a separable form for  $\boldsymbol{\zeta}$, $\dfrac{d{\xi}_n}{dn}$, $\dfrac{d \tau}{d n}$, with $\dfrac{d}{dn}=\boldsymbol{n}^T {\boldsymbol \nabla}$ and  $\xi_n=\boldsymbol{\xi}^T{\boldsymbol n}$.
Since $\dfrac{d{\xi}_n}{dn}$,   $\dfrac{d \tau}{d n}$ are independent, it implies :  \begin{equation*}
\left[\,b\,\right]=0\quad \text{and} \quad\left[b\, \boldsymbol{n}^T\boldsymbol{u}\right]=0
\end{equation*}
 Consequently from $[v] = [\boldsymbol n^T\boldsymbol{u}-D_n]=[\boldsymbol n^T\boldsymbol{u}]\neq 0$, we obtain : 
\begin{equation} 
b=0 
\label{Key4p}   
\end{equation}
Condition \eqref{Key4p} implies the continuity of all tangential derivatives of $b$ on $S_t$. Hence, the vector $\boldsymbol{\it \Theta}$ given by \eqref{Theta}   is identically null. The  relation $\delta a=0$ given by \eqref{key4} reduces to relations coming from the conservative form  \eqref{Motion equations} :
\begin{equation} 
\left[{\boldsymbol{N}^T}\left( \boldsymbol{J}\boldsymbol{K}%
^T-\mathcal A^T\frac{\partial \boldsymbol{J}}{\partial \boldsymbol{z}}+p\,  \boldsymbol{I}_{4}\right)  \right] =\boldsymbol 0^T\label{Key2}    
\end{equation}
They are supplemented by an additional relation  \eqref{Key4p}.  \\

 {\bf Remark:}  In the non-isentropic case (i.e. $\eta \ne const$) the  shock relations \eqref{Key2}  conserve  both the momentum and energy.  If we restrict our attention to  isentropic (or, more generally, barotropic) flows,   the conservation of   energy is not compatible with the conservation of momentum : `energy inequality' takes the place of `entropy inequality'.   This means that depending   on  physical situations associated with special fluid flows,  it may be necessary to consider   only  the space variations  $\boldsymbol{\xi}$ and  $\dfrac{d{\xi}_n}{dn}$, and not those  associated with the time variations.  Thus, the  number of shock conditions  may be less than  in the general case. In what follows,  we   study only the general situation. \\

 We   now express   conditions \eqref{Key4p} and \eqref{Key2} for the two special cases. 
 
\subsection{Capillary fluids}
For capillary fluids $b$ is denoted by $c$   (see \eqref{bcapillary}   in  Appendix \ref{Capillary fluids}) :  
\begin{equation*} 
c=\rho^2\,\boldsymbol {n}^T\dfrac{\partial\alpha}{\partial\left(\dfrac{\partial\rho}{\partial\boldsymbol x}\right)} \label{Key5p} 
\end{equation*} 
 In general, specific internal energy $\alpha$ is quadratic  in $\dfrac{\partial\rho}{\partial\boldsymbol x}$ :
\begin{equation*} 
\alpha= \beta (\rho, \eta)+\frac{1}{2}\,\lambda(\rho,  \eta)\,\dfrac{\partial\rho}{\partial\boldsymbol x}\left(\dfrac{\partial\rho}{\partial\boldsymbol x}\right)^T , \quad \lambda >0
\end{equation*} 
with given functions $\beta(\rho,\eta)$ and $\lambda(\rho, \eta)$.  Hence, on  the  shock :
\begin{equation} 
 \dfrac{\partial\rho}{\partial\boldsymbol x}\, \boldsymbol{n}  \equiv\frac{d\rho}{dn} =0 \label{gradrho}
\end{equation}
Relation \eqref{Key2} immediately yields :
\begin{equation*}
\left[\rho\,  v\,\boldsymbol u + 
p  \,\boldsymbol{n}  \right]=\boldsymbol{0}_3  \label{Key3p}
\end{equation*}
and
\begin{equation*}
\left[
e \,v  +p\,\boldsymbol{n}^T \boldsymbol{u}\right]=0 
\end{equation*}
which are  in the same  form as   the classical Rankine--Hugoniot relations for the momentum  and  energy, respectively. Let us remark  that $e$ and  $p$  depend here  on the first and second order space derivatives of $\rho$.

\subsection{Bubbly fluids}
 For bubbly fluids $b$ is denoted by $\kappa$ (see  \eqref{bbubbly} in Appendix \ref{Bubbly fluids}) :
\begin{equation*} 
 \kappa=\rho^2 v\, \dfrac{\partial w}{\partial{\dot \rho }} \label{Key5} 
\end{equation*} 
In general,  $w$ is quadratic in ${\dot \rho }$ :
\begin{equation*} 
 w= \beta (\rho,\eta)-\frac{1}{2}\,\lambda(\rho, \eta)\,{\dot \rho }^2, \quad\lambda >0
 \end{equation*} 
  with  given functions $\beta(\rho,\eta)$ and $\lambda(\rho, \eta)$.   On the shock \eqref{Key4p} becomes :
 \begin{equation} 
{\dot \rho }=0 \label {rhodot}
 \end{equation}
 Relation \eqref{Key2}  immediately yields :
 \begin{equation*}
 \left[\rho\,  v\,\boldsymbol u + 
 p  \,\boldsymbol{n}  \right]=\boldsymbol{0}_3  \label{Key3}
 \end{equation*}
  and
 \begin{equation*}
 \left[
  e\, v +p\,\boldsymbol{n}^T \boldsymbol{u}\right]=0 
 \end{equation*}
 which are the form of the classical Rankine--Hugoniot relations for the momentum  and energy, respectively, but with the pressure depending on the  second material  derivatives of $\rho$.

 \section{Conclusion}
We have obtained the Rankine--Hugoniot conditions in the cases where the internal specific energy depends on space and time  derivatives of density. As usually, these conditions express the conservation of mass, momentum and energy. Compared to the conventional conservation laws of the momentum and energy, they contain additional terms depending on the density derivatives.  Moreover, an additional relation \eqref{Key4p} to the classical jump conditions  is obtained   (see also \eqref{gradrho} and \eqref{rhodot}). 
 The meaning of \eqref{Key4p}  can easily be understood  in the case of capillary fluids. If we consider a rigid surface  in contact with a capillary fluid, the boundary condition   when the surface has no energy is in the form  \cite{Gouin-Kosinski} :
\begin{equation}
\frac{d\rho}{dn} =0.
\label{normal}
 \end{equation}
 Hence, this condition can be interpreted as the absence of 
 the interaction between  fluids separated by the shock front.  For bubbly fluids  the condition is 
\begin{equation}
 \dot\rho = 0.
 \label{material} 
 \end{equation}
This condition can be interpreted as the absence  on a moving  front of  the local kinetic energy which is proportional to $\dot\rho^2$. 

In both cases, the shock front has to be considered as a  geometrical surface without energy. Relations \eqref{normal}--\eqref{material} are the analogs of `balance of hyper momentum' appearing in elasticity \cite{Mindlin_1965,Truskinovsky_Zanzotto_1996}. An example of such a singular shock solution was found in \cite{Gavrilyuk_2020} in the case of dispersive shallow water equations which have the same mathematical  structure as the isentropic  equations of bubbly fluids.   At such a shock the condition  \eqref{material} was satisfied  in addition to the laws of conservation of mass and momentum. The energy equation played  in this case the role of `entropy inequality'.
 
Further studies are  needed, both  analytical and numerical, to understand better singular  shock solutions to dispersive systems of equations. \\
 
{\bf Acknowledgments :} The authors thank the anonymous referees for helpful suggestions. They  were partially supported by l'Agence Nationale de la Recherche, France (project SNIP ANR--19--ASTR--0016--01).

\appendix
 \section{Variation of the Jacobian}
 \label{Appendix 1}
Quadri-vector $ \displaystyle\boldsymbol{J}=\left(
	\begin{array}{c}
		\rho \\
		\rho\boldsymbol{u}
	\end{array}
	\right)$   is a  form of $\mathcal W$ of image ${\boldsymbol J}_0$ in $\mathcal W_0$. Then, \eqref{massconstr} can be rewritten in Lagrangian coordinates as :
\begin{equation*}
{\boldsymbol J} = \frac{\mathcal B}{\text{det}\,\mathcal B}\, {\boldsymbol J}_0(\boldsymbol{Z})\quad\text{with}\quad\mathcal{B}=\frac{\partial\boldsymbol z}{\partial\boldsymbol{Z}}\quad \text{and}\quad \text{Div}_0\,{\boldsymbol J}_0 =0  
\end{equation*}
Consequently :
\begin{equation*}
\delta\,{\boldsymbol J} = \left(\frac{\delta\,\mathcal  B}{\text{det}\,\mathcal B}-  \frac{1}{(\text{det}\mathcal{B})^2}\, \mathcal B\ \delta\,(\text{det}\,\mathcal B)\right)\,{\boldsymbol J}_0  
\end{equation*}
 By using the Euler-Jacobi identity :
\begin{equation*}   \delta\, (\det\,\mathcal B)\,  =\, \det \,\mathcal B  \ {\rm Tr}\, (\mathcal B^{-1}\, \delta\, \mathcal B )\quad\text{ with}\quad {\rm Tr}\, (\mathcal B^{-1}\, \delta\, \mathcal B )= {\rm Tr}\,\left(\frac{\partial\boldsymbol \zeta}{\partial z}\right) \equiv \text{Div}\, \boldsymbol\zeta
\end{equation*}
and we obtain :
\begin{equation*}
\delta\,{\boldsymbol J}  = \frac{1}{\text{det}\mathcal{B}}\left(\frac{\partial\boldsymbol\zeta}{\partial z}-( {\rm{Div}}\, { \boldsymbol \zeta })\,{\boldsymbol I}_4\right)\,\mathcal B\, {\boldsymbol J}_0 =  \left(\dfrac{\partial \boldsymbol\zeta} {\partial {\boldsymbol z}} \,
-\, ( {\rm{Div}}\, { \boldsymbol \zeta }) \,{\boldsymbol I}_4 \right)  {\boldsymbol J} 
\end{equation*}
{\section{Specific cases}
\label{Appendix 2}
\subsection{Capillary fluids}
\label{Capillary fluids}
One has from \eqref{formulae2}
\begin{equation*}
\boldsymbol{N}^T \mathcal A^T=\left(
-\rho\,\boldsymbol{n}^T\dfrac{\partial\alpha}{\partial\left(\dfrac{\partial\rho}{\partial\boldsymbol x}\right)},\ \boldsymbol{0}^T_3\right) 
\end{equation*}
Using  \eqref{moment} we  obtain :
\begin{equation*}
\boldsymbol {N}^T \mathcal  A^T\delta \boldsymbol {J} =   c \left(\text{div}\,\boldsymbol{\xi}-\frac{\partial \tau}{\partial \boldsymbol{x}}\boldsymbol{u}\right) 
\end{equation*}
 where : 
\begin{equation} 
 c =\rho^2\,\boldsymbol {n}^T\dfrac{\partial\alpha}{\partial\left(\dfrac{\partial\rho}{\partial\boldsymbol x}\right)} \label{bcapillary}
\end{equation} 
Since at the shock 
\begin{equation*}
\boldsymbol{N}^T  \boldsymbol{J }=\rho \left( \boldsymbol{n}^T 
\boldsymbol{u}- D_{n}\right)=\rho\,v 
\end{equation*}
we finally obtain :
\begin{equation*}
\boldsymbol{N}^T \mathcal A^T\frac{\partial \boldsymbol{J}}{\partial \boldsymbol{z}}= - \rho\,\boldsymbol{n}^T \dfrac{\partial\alpha}{\partial\left(\dfrac{\partial\rho}{\partial\boldsymbol x}\right)} \dfrac{\partial \rho }{\partial \boldsymbol{z}} 
\end{equation*}
and from \eqref{K^T} :
\begin{equation*}
\boldsymbol{N}^T \boldsymbol{J}\boldsymbol{K}^T=  \rho\, v\left( -\frac{1}{2}\left|\boldsymbol{u}\right|^2-\alpha-\rho\,\dfrac{\partial \alpha}{\partial\rho
}+\text{div}\left\{\rho\,\dfrac{\partial\alpha}{\partial\left(\dfrac{\partial\rho}{\partial\boldsymbol x}\right)}\right\},\ 
\boldsymbol{u}^T \right)   
\end{equation*}

\subsection{Bubbly fluids}
\label{Bubbly fluids}
 From Eqs. \eqref{moment} and \eqref{formulae} we   obtain :
\begin{equation*}
\boldsymbol {N}^T \mathcal A^T\delta \boldsymbol {J}=\kappa  \left(\text{div}\,\boldsymbol{\xi}-\frac{\partial \tau}{\partial \boldsymbol{x}}\boldsymbol{u}\right)  
\end{equation*}
 where : 
\begin{equation}
 \kappa=\rho^2 v\, \dfrac{\partial w}{\partial\overset{\centerdot }{\rho }}\label{bbubbly}
\end{equation}
Since
 \begin{equation*}
 \boldsymbol{N}^T  \boldsymbol{J }=\rho \left( \boldsymbol{n}^T\
 \boldsymbol{u}- D_{n}\right)=\rho\,v ,\quad \boldsymbol{N}^T \mathcal A^T={\left( -\rho\,v\dfrac{\partial w}{\partial \overset{\centerdot }{\rho }},\ \boldsymbol{0}^T_3\right)} 
 \end{equation*}
 one has    
 \begin{equation*}
 \boldsymbol{N}^T \mathcal A^T\frac{\partial \boldsymbol{J}}{\partial \boldsymbol{z}}= - \rho v \dfrac{\partial w}{\partial \overset{\centerdot }{\rho }}\dfrac{\partial \rho }{\partial 	\boldsymbol{z}} 
 \end{equation*}
and  
 \begin{equation*}
 \boldsymbol{N}^T \boldsymbol{J}\boldsymbol{K}^T=  \rho\, v\left( -\frac{1}{2}\left|\boldsymbol{u}\right|^2-w-\rho\,\dfrac{\partial w}{\partial\rho
 }+ \dfrac{\partial w}{\partial
 	\overset{\centerdot }{\rho }}  \frac{\partial\rho}{\partial \boldsymbol{x}}\, \boldsymbol{u}+\rho\,\dfrac{d }{dt}\left( \dfrac{\partial w}{%
 	\partial \overset{\centerdot }{\rho }}\right) ,\ 
 \boldsymbol{u}^T- \dfrac{\partial w}{\partial \overset{%
 		\centerdot }{\rho }}\frac{\partial\rho}{\partial \boldsymbol{x}} \right)
 \end{equation*}}
\section{Technical lemmas}
\label{Technical lemmas}

{\it For any scalar field $b$ of the $\textit{3-D}$ physical space, we have the property :
	\begin{equation*}
	b\,  {\rm div}\, \boldsymbol{\xi}=\boldsymbol {n}^T\, {\rm rot}\,\left(\boldsymbol {n}\times b\,\boldsymbol{\xi}\right)+ b\,\frac{d\xi_n}{dn}- b\,\boldsymbol{n}^T \left(\frac{\partial\boldsymbol{n}}{\partial\boldsymbol{x}}\right)^T \boldsymbol{\xi}-b\, H \boldsymbol{n}^T \boldsymbol{\xi} +\frac{\partial b}{\partial
		\boldsymbol{x}}\left(\boldsymbol{n}\boldsymbol{n}^T-\boldsymbol{I}\right)\, \boldsymbol{\xi} 
	\end{equation*}
	where $H=-{\rm div}\, {\boldsymbol n}$ is the sum of principal curvatures of  $S_t$, $\xi_n ={\boldsymbol n}^T {\boldsymbol \xi}$, and $\displaystyle\frac{d}{dn}$ is the normal derivative to $S_t$.}  Here one supposes that the normal vector field $\boldsymbol{n}$ is locally extended in the vicinity of $S_t$. \\

\begin{proof}	
Let $\boldsymbol{a}$ be a unit vector field, and $\boldsymbol{b}$ be any vector field. Then   
		\begin{equation}
		{\rm div}\, \boldsymbol{b}=\boldsymbol{a}^T\, {\rm rot}\,\left(\boldsymbol {a}\times \boldsymbol b\right)+ \boldsymbol{a}^T\boldsymbol{b}\, {\rm div}\, \boldsymbol{a}+ \boldsymbol{a}^T\dfrac{\partial \boldsymbol{b}}{\partial \boldsymbol{x}}\, \boldsymbol{a} 
		\label{div}
		\end{equation}
To obtain the result, it is sufficient to multiply  the  identity 
		\begin{equation*}
		{\rm rot}\,\left(\boldsymbol {a}\times \boldsymbol b\right)=\boldsymbol {a}\,  {\rm div}\, \boldsymbol{b}-\boldsymbol {b}\,  {\rm div}\, \boldsymbol{a}+\dfrac{\partial \boldsymbol{a}}{\partial \boldsymbol{x}}\,\boldsymbol {b}-\dfrac{\partial \boldsymbol{b}}{\partial \boldsymbol{x}}\,\boldsymbol {a}
		\end{equation*} by $\boldsymbol{a}^T$. 	
		Also, one has the property :
		\begin{equation}
		\boldsymbol{n}^T\,\left(\frac{\partial \boldsymbol b}{\partial \boldsymbol x}\right)\,\boldsymbol{n}= \frac{db_n}{dn}-\,\boldsymbol{n}^T \left(\frac{\partial\boldsymbol{n}}{\partial\boldsymbol{x}}\right)^T \boldsymbol{b},
		\label{normal derivative}
		\end{equation}
		with $b_n={\boldsymbol n}^T\boldsymbol b$. 
	 We take  $\boldsymbol {a}=\boldsymbol {n}$ and $\boldsymbol {b}=b\,\boldsymbol {\xi}$ in \eqref{div}. 
		One obtains 
	\begin{equation*}
	\text {div}  \, (b\,\boldsymbol \xi) = \boldsymbol{n}^T\, \text {rot}\left(\boldsymbol{n}\times b\, \boldsymbol \xi\right)-b\, H \boldsymbol{n}^T \boldsymbol \xi +\boldsymbol{n}^T\,\left(\frac{\partial b\,\boldsymbol \xi}{\partial \boldsymbol x}\right)\, \boldsymbol{n}. 
	\end{equation*}
The property \eqref{normal derivative} allows us to complete the proof.  
\end{proof}	\\
\\

{\it 	For any scalar field $b$ of the $\textit{3-D}$ physical space, we have the property :
	\begin{equation*}
	b\, \frac{\partial \tau}{\partial \boldsymbol{x}}\boldsymbol{u}=\boldsymbol {n}^T\, {\rm rot}\,\left(\boldsymbol {n}\times  b\,\tau\, \boldsymbol{u}\right) -\left(b\, H \boldsymbol{n}^T \boldsymbol{u} -\boldsymbol{n}^T\frac{\partial b\,\boldsymbol{u} }{\partial\boldsymbol{x} }\,\boldsymbol{n}+ \rm{ div}\,(b\boldsymbol{u})\right) \tau+b\,\boldsymbol{n}^T \boldsymbol{u}\,\frac{d\tau}{dn} 
	\end{equation*}}

\begin{proof} 
	From relations
	\begin{equation*}
	b\, \frac{\partial \tau}{\partial \boldsymbol x}\,\boldsymbol{u}= \text{div}\left(b\,\tau\, \boldsymbol{u}\right)-
	\tau\,\text{div}\left(b\,  \boldsymbol{u}\right) 
	\end{equation*}
and
\begin{equation*}
\text{div}\left(b\,\tau\, \boldsymbol{u}\right) = \boldsymbol {n}^T\, {\rm rot}\,\left(\boldsymbol {n}\times b\,\tau\,\boldsymbol{u}\right)-b\, H \tau\, \boldsymbol{n}^T \boldsymbol u+\tau\,\boldsymbol{n}^T\,\frac{\partial b\,\boldsymbol u}{\partial \boldsymbol x}\, \boldsymbol{n}
	+b\,\boldsymbol{n}^T \boldsymbol u\,\frac{\partial \tau}{\partial \boldsymbol x}\, \boldsymbol n 
	\end{equation*}
 we deduce  the relation.
\end{proof}\\

\end{document}